\documentclass[preprints, article, accept, pdftex, moreauthors, letterpaper]{Definitions/mdpi} 

\firstpage{1} 
\pubvolume{1}
\issuenum{1}
\articlenumber{0}
\pubyear{2023}
\copyrightyear{2023}
\datereceived{} 
\dateaccepted{} 
\datepublished{} 
\hreflink{https://doi.org/} 
\pdfoutput=1

\usepackage{nameref,hyperref}
\hypersetup{
    colorlinks=true,
    linkcolor=blue,
    filecolor=magenta,      
    urlcolor=blue,
    pdfpagemode=FullScreen,
}

\newcommand{\ket}[1]{\left| #1 \right>} 
\newcommand{\bra}[1]{\left< #1 \right|} 
\newcommand{\braket}[2]{\langle #1 \vphantom{#2} | #2 \vphantom{#1} \rangle} 
\let\baraccent=\= 
\renewcommand{\=}[1]{\stackrel{#1}{=}} 

\newcommand{\ID}{\text{ID}}
\newcommand{\QID}{\text{QID}}
\newcommand{\KLD}{\text{KLD}}
\newcommand{\Tr}{\text{Tr}}
\renewcommand{\P}{\text{Pr}} 
\renewcommand{\H}{\mathcal{H}} 
\newcommand{\T}{\mathcal{T}} 
\newcommand{\p}{\theta} 
\newcommand{\Part}{\Theta}
\renewcommand{\O}{\emptyset}
\newcommand{\powerset}{\mathbb{P}}
\newcommand{\rmm}{\rho_{mm}}
\DeclareMathOperator{\ii}{\mathit{ii}} 

\renewcommand{\a}{\alpha}

\newcommand*{\argmin}{\operatornamewithlimits{argmin}\limits}
\newcommand*{\argmax}{\operatornamewithlimits{argmax}\limits}

\Title{Measuring the integrated information of a quantum mechanism}

\TitleCitation{Measuring the integrated information of a quantum mechanism}

\Author{Larissa Albantakis $^{1,4}$*\orcidA{}, Robert Prentner $^{2,4}$\orcidB{} and Ian Durham $^{3,4}$\orcidC{} }

\AuthorNames{Larissa Albantakis, Robert Prentner and Ian Durham}

\AuthorCitation{Albantakis, L.; Prentner, R.; Durham, I.}

\address{%
$^{1}$ \quad Department of Psychiatry, University of Wisconsin–Madison, Madison, WI 53719, USA\\
$^{2}$ \quad Munich Center for Mathematical Philosophy, Ludwig-Maximilians-University Munich, 80539, Germany; robert.prentner@amcs.science\\
$^{3}$ \quad Department of Physics, Saint Anselm College, Manchester, NH 03102, USA ; idurham@anselm.edu\\
$^{4}$ \quad Association for Mathematical Consciousness Science, Munich, Germany}

\corres{Correspondence: albantakis@wisc.edu}

\abstract{Originally conceived as a theory of consciousness, integrated information theory (IIT) provides a theoretical framework intended to characterize the compositional causal information that a system, in its current state, specifies about itself. 
However, it remains to be determined whether IIT as a theory of consciousness is compatible with quantum mechanics as a theory of microphysics. Here, we present an extension of IIT's latest formalism to evaluate the mechanism integrated information (\texorpdfstring{$\varphi$})) of a system subset to finite-dimensional quantum systems (e.g., quantum logic gates). To that end, we translate a recently developed, unique measure of intrinsic information into a density matrix formulation, and extend the notion of conditional independence
to accommodate quantum entanglement. 
The compositional nature of the IIT analysis might shed some light on the internal structure of composite quantum states and operators that cannot be obtained using standard information-theoretical analysis.
Finally, our results should inform theoretical arguments about the link between consciousness, causation, and physics from the classical to the quantum.}


\begin{document}
\section{Introduction}

Integrated information theory \cite{Oizumi2014, Tononi2016, Albantakis2020, Albantakis2022_4_0} stands out as one theory of consciousness that explicitly proposes a formal framework for identifying conscious systems. 
Specifically, IIT provides requirements about the intrinsic causal structure of a system that supports consciousness, based on the essential (``phenomenal'') properties of experience.
Its formal framework evaluates the causal powers that a set of interacting physical units exerts on itself in a compositional manner \cite{Oizumi2014, Albantakis2019cc, Barbosa2021, Grasso2021, Albantakis2022_4_0}. 

IIT does not presuppose that consciousness arises at the level of neurons rather than atoms, molecules, or larger brain areas, but assumes causation to be a central concept for analyzing a physical system across the hierarchy from the microphysical to the macroscopic \cite{Hoel2013, Hoel2016, Tononi2016, Marshall2018, Albantakis2021}.
One prediction of IIT is that consciousness appears at the level of organization at which the intrinsic causal powers of a system are maximized \cite{Tononi2016}.
To that end, IIT offers a formal framework for causal emergence that compares the amount of integrated information of macroscopic causal models to their underlying microscopic system descriptions \cite{Hoel2016, Marshall2018}.  
Whether causality plays a fundamental role in physics, in particular in quantum physics, is, however, contested \cite{Brukner2014,DAriano2018}. 

IIT's causal framework has been formalized for discrete dynamical systems in the classical realm \cite{Oizumi2014, Albantakis2015, Mayner2018, Gomez2021, Albantakis2022_4_0}. 
Accordingly, in prior studies \cite{Hoel2013, Hoel2016, Marshall2018}, micro-level systems corresponded to classical causal networks \cite{Pearl2000, Albantakis2019}, constituted of individual, conditionally independent physical units, that can (in principle) be manipulated and whose states can be observed. 
Thus, it remains to be determined whether IIT is compatible with quantum mechanics \cite{Barrett2019, Carroll2021}.

Here, we are interested in the question of whether it is possible to apply or extend the causal framework of IIT to quantum mechanics, starting with IIT's measure of mechanism integrated information ($\varphi$) \cite{Barbosa2021, Albantakis2022_4_0}. Several attempts to apply the general principles of IIT to quantum systems have recently been proposed \cite{Tegmark2015, Zanardi2018, Kleiner2021}. Of these, the work by Zanardi et al. \cite{Zanardi2018} comes closest to a direct translation of the previous version of the theory (``IIT 3.0'') \cite{Oizumi2014} into a quantum-mechanical framework. However, this translation is not unique, does not converge to the classical formalism for essentially classical state updates, and also does not explicitly take the philosophical grounding of IIT as a theory of consciousness into account. 

Our objective is to accurately transform the various steps of the IIT formalism in its latest iteration (``IIT 4.0'') \cite{Barbosa2021, Albantakis2022_4_0} to be applicable to both classical and quantum systems. As a first step, here we propose an extension of the IIT formalism to evaluate the integrated information ($\varphi$) of a mechanism within a system \cite{Barbosa2021} to quantum mechanisms (e.g. quantum logic gates). To enable a direct quantitative comparison, quantum integrated information should converge to the classical formulation if the quantum system under consideration has a classical analog. 
Our main contributions, of merit beyond the scope of IIT, are (1) the translation of a newly defined, unique measure of intrinsic information \cite{Barbosa2020, Barbosa2021} to a quantum density matrix formalism, and (2) a formulation of the causal constraints specified by a partial quantum state. To that end, we extend the notion of conditional independence and causal marginalization \cite{Albantakis2019} to accommodate quantum entanglement. 
In the results section, we will apply our theoretical developments to classical computational gates and their quantum analogues (such as the CNOT gate), as well as quantum states and gates without a classical counterpart. 
The additional challenges of evaluating the integrated information of an entire quantum system will be outlined in the discussion. 


While our investigation is based on IIT's formal framework, it raises questions that apply to any theory of consciousness and its relation to (micro) physics \cite{Tegmark2015,sep-quantum-consciousness,Prakash20}. However, we also want to emphasize that this work is not concerned with the question of whether biological systems (in particular the brain) should be treated quantum-theoretically or classically. The question of whether a theory of consciousness is generally applicable across microscopic and macroscopic scales and thus consistent with our knowledge of micro physics is important in either case. 
Our work is also not directly related to the potential role of consciousness in quantum measurements and the operational collapse of the wave function \cite{wigner1995remarks, Kremnizer2015, Chalmers2021}, although we briefly discuss several difficulties in applying IIT's causal analysis to measurement dynamics. 
At the very least, our results should inform theoretical arguments about the link between consciousness, causation, and physics from the classical to the quantum \cite{Atmanspacher22}. Finally, the compositional nature of the IIT analysis might also shed some light on the internal structure of composite quantum states and operators that cannot be obtained using standard information-theoretical analysis. To that end, we provide python code to analyze quantum mechanisms of two and three qubits, available at \url{https://github.com/Albantakis/QIIT} (accessed on 30 December 2022).

\section{Theory}


The purpose of IIT's formal analysis is to evaluate the irreducible causal information that a system in a particular state specifies about itself. Notably, IIT's notion of causal information differs from other information-theoretical measures in multiple ways: it is intrinsic (evaluated from the perspective of a mechanism within the system), state-dependent (evaluated for particular states, not state averages), causal (evaluated against all possible counterfactuals of a system transition \cite{Pearl2000, Albantakis2019}), and irreducible (evaluated against a partition of the mechanism into independent parts). 
Moreover, the IIT analysis is \textit{compositional} \cite{Albantakis2019cc}: instead of only analyzing the system as a whole, or only its elementary components, any system subset counts as a candidate \textit{mechanism} that may specify its own irreducible cause and effect within the system. The IIT analysis thus evaluates the irreducible cause-effect information ($\varphi$) of every subset of units within the system \cite{Barbosa2021}, which amounts to ``unfolding'' the system's cause-effect structure.

In the following, we will extend IIT's $\varphi$-measure, the integrated information of a mechanism, to be applicable to finite-dimensional quantum systems. While the full IIT analysis assumes a dynamical system of interacting units, mechanism integrated information ($\varphi$) can be evaluated in a straightforward manner for any type of input-output logic, such as sets of logic gates, or whole computational circuits, as well as information channels (see Figure \ref{fig:cCNOT} as an example). 
For a classical template of our quantum version of mechanism integrated information ($\varphi$) we follow Barbosa et al. \cite{Barbosa2021}, including minor updates within the most recent formulation ``IIT 4.0'' \cite{Albantakis2022_4_0}, which is briefly reviewed in the following. 
As a result, the quantum integrated information of a mechanism as defined below coincides with the classical measure \cite{Barbosa2021, Albantakis2022_4_0} if the quantum system under consideration has a classical analog.

\subsection{Classical Systems}
In the canonical IIT formalism, a classical physical system $S$ of $n$ interacting units is defined as a stochastic system $S = \{S_1, S_2, \ldots, S_n\}$ with finite, discrete state space $\Omega_S = \prod_i \Omega_{S_i}$ and current state $s_t \in \Omega_S$ \cite{Barbosa2021} that evolves according to a transition probability function 

\begin{equation}
\label{eq:tpm}
\T_S \equiv p(s_{t+1}\mid s_t) = \P(S_{t + 1} = s_{t+1}\mid S_t = s_t), \quad s_t, s_{t+1} \in \Omega_S,
\end{equation}
with the additional requirement that $S$ corresponds to a causal network \cite{Barbosa2021}. This implies that the conditional probabilities $p(s_{t+1}|s_t)$ are well-defined for all possible states 

\begin{equation}
    \exists \, p(s_{t+1}|s_t) \, \forall \, s_{t}, s_{t+1} \in \Omega_S,
\end{equation}
with $p(s_{t+1}|s_t) = p(s_{t+1}|do(s_t))$ \cite{Albantakis2019, Janzing2013, Ay2008, Pearl2000}, where the ``do-operator'' $do(s_t)$ indicates that $s_t$ is imposed by intervention. Moreover, the individual random variables $S_i \in S$ are assumed to be conditionally independent from each other given the preceding state of $S$, 

\begin{equation}
\label{eq:condind}
p(s_{t+1}\mid s_t) = \prod_{i=1}^n p(s_{i,t+1}|s_t),
\end{equation}
which has to be revisited in the quantum case. If $S$ is an open system within a larger universe $U$ with current state $u_t \in \Omega_U$, variables $W = U \setminus S$ are treated as fixed background conditions throughout the causal analysis \cite{Oizumi2014, Barbosa2021}.  

\begin{figure}[p!]
\includegraphics[width=\textwidth]{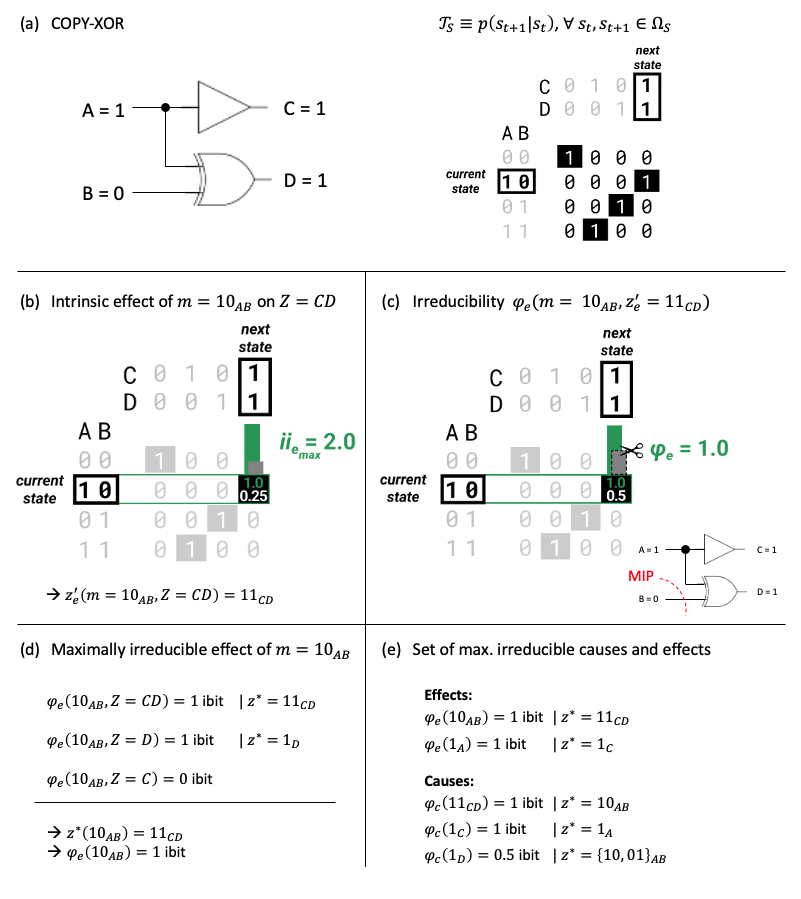}
\caption{Outline of the IIT analysis applied to a classical COPY-XOR gate. \textbf{(a)} The COPY-XOR gate and its (deterministic) transition probability function $\T_S$ depicted by a probability matrix. Unit $C$ is a copy of the input bit $A$, and $D$ corresponds to an XOR function of both input bits ($A,B$). For input state $AB = 10$ (also denoted $10_{AB}$) the COPY-XOR gate outputs $CD = 11$ (denoted $11_{CD}$). 
\textbf{(b)} Based on $\T_S$, we can identify the intrinsic effect of a mechanism $M$ in its current state $m$ over a purview $Z$ as the effect state $z'_e$ with maximal intrinsic effect information $ii_e$. For $m= 10_{AB}$ and $Z=CD$, the intrinsic effect is $z'_e = 11_{CD}$. \textbf{(c)} Next, we assess the irreducibility of the intrinsic effect by computing the integrated information $\varphi_e(m,Z)$ over the minimum partition (MIP).
\textbf{(d)} To identify the maximally irreducible effect of a mechanism $m$, we compare $\varphi_e(m,Z)$ across all possible effect purviews $Z$. Here, the maximally irreducible effect of $m = 10_{AB}$ is $z^*_e = 11_{CD}$, because it specifies a maximum of $\varphi_e$ and is the largest purview that does so (see text for details). \textbf{(e)} For a given system, we identify all maximally irreducible causes and effects. Given the input state $AB = 10$, the classical IIT analysis identifies two irreducible effects, the first-order mechanism $1_A$ specifies the effect $1_C$, and the second-order mechanism $10_{AB}$ specifies the effect $11_{CD}$. Given the output state $CD = 11$, the IIT analysis identifies three irreducible mechanisms, including the mechanism $1_D$ with purview $10_{AB}$ or $01_{AB}$ (which are tied). Both intrinsic information ($ii$) and integrated information ($\varphi$) are quantified in ``ibit'' units (see text below).}
\label{fig:cCNOT}
\end{figure}

A mechanism $M \subseteq S$ is a subset of the system $S$ with current state $m_t \in \Omega_M$. 
The intrinsic information that a mechanism $M$ in state $m_t$ specifies over a ``purview'' $Z_{t\pm1} \subseteq S$, is defined by a difference measure $\ii(m_t, Z_{t\pm1})$, which quantifies how much $m_t$ constrains the state of $Z_{t\pm1}$ compared to chance, but also takes its \textit{selectivity} into account (how much the mechanism specifies a particular state of $Z_{t\pm1}$) \cite{Barbosa2021, Albantakis2022_4_0}. 
The mechanism's integrated information $\varphi(m_t, Z, \theta)$ is then evaluated over the maximal cause and effect states $z'_{c/e}$ identified by the intrinsic information measure. It quantifies how much the mechanism $m_t$ constrains $z'_{c/e}$ as \textit{one} mechanism, compared to a partition $\p$ 

\begin{equation}
\label{eq:parts}
\p = \{ (M^{(1)}, Z^{(1)}), (M^{(2)}, Z^{(2)}), \ldots, (M^{(k)}, Z^{(k)}) \},
\end{equation}
of the mechanism and purview into $k$ independent parts \cite{Oizumi2014, Barbosa2021}. 
Below we will define all relevant quantities for computing the mechanism integrated information $\varphi(m_t)$ following Barbosa et al. \cite{Barbosa2021} with minor updates from \cite{Albantakis2022_4_0}. Figure \ref{fig:cCNOT} outlines the steps of IIT's causal analysis for a simple example system, a COPY-XOR gate.

\subsubsection{Cause and effect repertoires}
How the state of a mechanism $M = m$ constrains the possible states of a purview $Z$ is captured by a product probability distribution $\pi(Z|m)$, which can be computed from the system's transition probability function (Eqn.~\ref{eq:tpm}) \cite{Oizumi2014, Albantakis2019, Barbosa2021}. Specifically, $\pi_c(Z|m) = \pi(Z_{t-1}|m_t)$ is the ``cause repertoire'' of $m$ over $Z$, and $\pi_e(Z|m) = \pi(Z_{t+1}|m_t)$ is the ``effect repertoire''. 
Without lack of generality, in what follows we will focus on effects of $m_t$ on purviews $Z = Z_{t+1}$ and omit update indices ($t-1$, $t$, $t+1$) unless necessary. 

\begin{figure}[h!]
\includegraphics[width=\columnwidth]{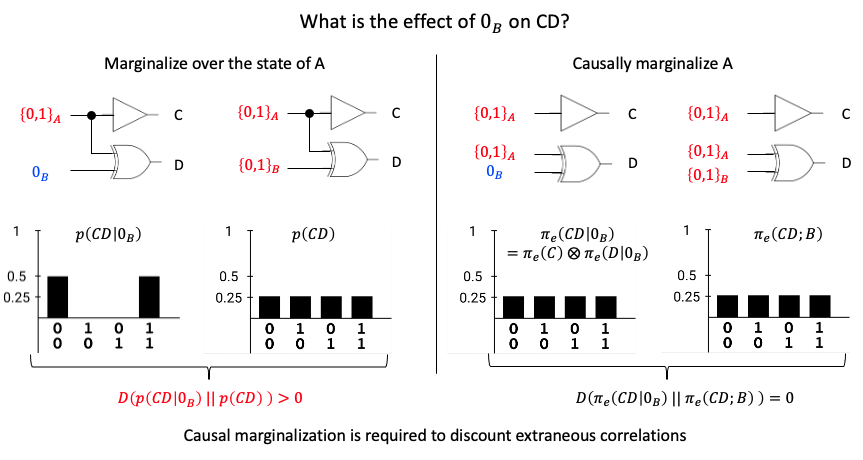}
\caption{Causal marginalization. Let us assume we want to identify the effect of the input bit $B = 0$ (or $0_B$) on the output $CD$ in the COPY-XOR system of Figure \ref{fig:cCNOT}. Intuitively, by itself, $0_B$ does not have an effect on $C$, as it does not input into $C$. It also has no effect on $D$, because, by itself, it specifies no information about the output state of the XOR $D$.
However, simply marginalizing the input $A$ (averaging over all possible input states of $A$, while maintaining the common inputs from $A$ to $C$ and $D$) would result in a ``spurious'' correlation between the output bits that is not due to $B$, but instead due to the common inputs from $A$. 
Capturing the fact that $0_{B}$ by itself has no effect on $CD$ requires causal marginalization (independent marginal inputs to each unit in the effect purview).}
\label{fig:cmarg}
\end{figure}

To capture the constraints on $Z$ that are due to the mechanism in its state ($M = m$) and nothing else, it is important to remove any contributions to the repertoire from outside the mechanism. This is done by ``causally marginalizing'' all variables in $X = S\setminus M$ \cite{Oizumi2014, Albantakis2019, Barbosa2021}. When evaluating the constraints of $m$ onto a single unit $Z_i \in Z$, causal marginalization amounts to imposing a uniform distribution as $p(X_t)$. 
The effect repertoire of a single unit $Z_i \in Z$ is thus defined as

\begin{equation}
\label{eq:cmarg}
\pi_e(Z_i \mid m) = |\Omega_{X}|^{-1} \sum_{x_t \in \Omega_{X}} p\left(Z_{i, t+1} \mid m_t,x_t\right).
\end{equation}
In the general case of an effect repertoire over a set $Z$ of $|Z|$ units (where $|Z|$ denotes the cardinality of the set of units $Z$), each $Z_i \in Z$ must receive independent inputs from units in $X = S\setminus M$ to discount correlations from units in $X$ with divergent outputs to multiple units in $Z$ (see Figure~\ref{fig:cmarg}). Formally, this amounts to using product probabilities $\pi(Z|m)$ instead of standard conditional probabilities $p(Z|m)$ (again imposing a uniform interventional distribution).
The effect repertoire over a set $Z$ of $|Z|$ units $Z_i$ is thus defined as the product of the effect repertoires over individual units 

\begin{equation}
\label{eq:effect_repertoire}
\pi_e(Z \mid m) = \bigotimes_{i=1}^{|Z|} \pi_e(Z_i \mid m),
\end{equation}
where $\bigotimes$ is the Kronecker product of the probability distributions. As in \cite{Albantakis2022_4_0}, we define the unconstrained effect repertoire as the marginal distribution

\begin{equation}
\label{eq:UCE}
    \pi_e(Z;M) = |\Omega_{M}|^{-1} \sum_{m \in \Omega_{M}} \pi_e(Z \mid m).
\end{equation}
The cause repertoire $\pi_c(Z|m)$ is obtained using Bayes' rule over the product distributions of the corresponding effect repertoire (for details see \cite{Barbosa2021, Albantakis2022_4_0}). The unconstrained cause repertoire $\pi_c(Z)$ is simply the uniform distribution over the states of $Z$.

\subsubsection{Intrinsic difference (ID)}

The classical version of mechanism integrated information ($\varphi$) evaluates the difference between two probability distributions $P = [p_1,...,p_N]$ and $Q = [q_1,...,q_N]$ based on a newly developed information measure, the ``intrinsic difference'' (ID) \cite{Barbosa2020, Barbosa2021}. 
The ID measure is uniquely defined based on three desired properties: \textit{causality}, \textit{specificity}, and \textit{intrinsicality}, which align with the postulates of IIT \cite{Barbosa2020, Barbosa2021, Albantakis2022_4_0}. Specifically, 

\begin{equation}
\label{eq:ID}
\ID(P,Q) = \max_\a \left( p_\a \log \left( \frac{p_{\a}}{q_{\a}} \right) \right),
\end{equation}
where $\a$ denotes a particular state in the distribution.

Formally, the ID is related to the Kullback-Leibler Divergence (KLD) or ``relative entropy'' measure, 

\begin{equation}
\label{eq:KLD}
\KLD(P,Q) = \sum_\a p_\a \log \left( \frac{p_{\a}}{q_{\a}} \right).
\end{equation}
While the KLD can be viewed as an average of the point-wise mutual information $\log \left( \frac{p_{\a}}{q_{\a}} \right)$ across states, the ID is instead defined based on the state that maximizes the difference between distributions (specificity property). For fully selective distributions (there is one state with probability one), the ID thus coincides with the KLD and is additive. Otherwise, the ID is subadditive and decreases with indeterminism (intrinsicality property). 
As argued in \cite{Barbosa2021}, this allows the ID to capture the information specified by a mechanism within a particular system. 
From the perspective of a mechanism the system has to be taken \textit{as is} (intrinsic perspective), while the KLD evaluates information from the perspective of a channel designer with the possibility to perform error correction (extrinsic perspective) \cite{Barbosa2020}.
To highlight this difference, the unit assigned to the ID measure is labeled an ``ibit'' or ``intrinsic bit''. Logarithms are evaluated with base 2 throughout. Formally, the ``ibit'' corresponds to a point-wise information value measured in bits weighted by a probability.

\subsubsection{Identifying intrinsic causes and effects}
Based on the intrinsic difference \eqref{eq:ID}, the intrinsic effect information that the mechanism $M = m$ specifies over a purview $Z$ can be quantified by comparing its effect repertoire $\pi_e(Z|m)$ to chance, that is, to the unconstrained effect repertoire $\pi_e(Z;M)$ \eqref{eq:UCE},

\begin{equation}
\label{eq:ii}
    \ii_e(m, Z) = \ID \left(\pi_e(Z|m), \pi_e(Z;M)\right)
\end{equation}

The specific state $z'_e \in \Omega_Z$ over which \eqref{eq:ii} is maximized corresponds to the intrinsic effect of the mechanism $M = m$ on the purview $Z$,

\begin{equation}
    z'_e(m, Z) = \argmax_{z \in \Omega_Z} \left(\pi_e(Z|m) \log \left( \frac{\pi_e(Z|m)}{\pi_e(Z;M)} \right) \right).
\end{equation}
The intrinsic cause $z'_c(m,Z)$ is defined in the same way based on the respective cause repertoires. (Note that the definition of the intrinsic cause information $\ii_c$ and, consequently, also the integrated cause information $\varphi_c$, has been updated in \cite{Albantakis2022_4_0} compared to \cite{Barbosa2021}. However, this update of the classical formulation is of no consequence in the quantum case and is thus not further discussed herein.)

\subsubsection{Disintegrating partitions}
The integrated effect information $\varphi_e(m,Z, \theta)$ quantifies how much the mechanism $m$ specifies the intrinsic effect $z'_e(m,Z)$ as \textit{one} mechanism and is assessed by comparing the effect probability $\pi(z'_e\mid m)$ to a partitioned effect probability $\pi^\p_e(z'_e\mid m)$ in which certain connections from $M$ to $Z$ are severed (causally marginalized).

Barbosa et al. \cite{Barbosa2021} (see also \cite{Albantakis2019, Albantakis2022_4_0}) define the set of possible partitions $\p \in \Part(M,Z)$ as

\begin{adjustwidth}{-0.75in}{0in}
\begin{multline}
\label{eq:partitions}
\Part(M,Z) = \Bigg\{ \{(M^{(i)}, Z^{(i)})\}_{i=1}^{k} \Bigg\vert k \in \{2,3,4,\ldots\}, M^{(i)} \in \powerset(M), Z^{(i)} \in \powerset(Z), 
\\ 
\bigcup M^{(i)} = M, \bigcup Z^{(i)} = Z, 
Z^{(i)} \cap Z^{(j)} = M^{(i)} \cap M^{(j)} = \O \; \forall \; i \neq j, M^{(i)} = M \Longrightarrow Z^{(i)} = \O\Bigg\}.
\end{multline}
\end{adjustwidth}
In words, for each $\p \in \Part(M,Z)$ it holds that $\{M^{(i)}\}$ is a partition of $M$ and $\{Z^{(i)}\}$ is a partition of $Z$ (as indicated in Eqn. \eqref{eq:parts}), but the empty set may also be used as a part ($\powerset$ denotes the powerset). However, if the whole mechanism is one part ($M^{(i)} = M$), then it must be cut away from the entire purview. This definition guarantees that any $\p \in \Part(M,Z)$ is a ``disintegrating partition'' of $\{M,Z\}$: it either ``cuts'' the mechanism into at least two independent parts if $|M| > 1$, or it severs all connections between $M$ and $Z$, which is always the case if $|M| = 1$, where again $|M|$ denotes the cardinality of the set of units $M$.

Given a partition $\p \in \Part(M,Z)$ constituted of $k$ parts (see Eq.~\eqref{eq:partitions}), we can define the partitioned effect repertoire

\begin{equation}
\label{eq:partitioned_effect}
\pi_e^\p(Z \mid m) = \bigotimes_{i = 1}^k \pi_e(Z^{(i)} \mid m^{(i)}), 
\end{equation}
with $\pi(\varnothing|m^{(i)}) = \pi(\varnothing) = 1$. In the case of $m^{(i)} = \varnothing$,  $\pi_e(Z^{(i)}|\varnothing)$ corresponds to the fully partitioned effect repertoire

\begin{equation}
\label{eq:FPCE}
    \pi_e(Z\mid \varnothing) = \bigotimes_{i=1}^{|Z|} \sum_{s_{t} \in \Omega_{S}} p(Z_{i, t+1} \mid s_{t}) |\Omega_{S}|^{-1}.
\end{equation}

\subsubsection{Mechanism integrated information}

In all, the general form of $\varphi_e(m,Z,\p)$ corresponds to that of the intrinsic difference $\ID$ \eqref{eq:ID}, albeit over the specific effect state $z'_e$

\begin{equation}
\label{eq:phi}
    \varphi_e(m,Z,\theta) = \varphi_e(m, z'_e, \theta) = \pi_e(z'_e \mid m) \log \left( \frac{\pi_e(z'_e \mid m)}{\pi_e^\theta(z'_e \mid m)} \right).
\end{equation}

Quantifying the integrated effect information of a mechanism $m_t$ within a system $S$, moreover requires an optimization across all possible partitions $\p \in \Part$ to identify the minimum partition (MIP) 

\begin{equation}
\label{eq:mip}
\p' = \argmin_{\p \in \Theta(M,Z)} \frac{\varphi_e(m,Z,\theta)}{\displaystyle\max_{\T'_S} \varphi_e(m,Z,\theta)}.
\end{equation}
The normalization factor $\max_{\T'_S} \varphi_e(m,Z,\theta)$ ensures that the minimum partition is evaluated against its maximum possible value across all possible system $\T'_S$ of the same dimensions as the original system. It was introduced in \cite{Albantakis2022_4_0} and shown to correspond to the number of possible pairwise interactions affected by the partition.

The integrated effect information of a mechanism over a particular purview $Z$ then corresponds to $\varphi_e(m, Z) = \varphi_e(m, Z, \p')$ (which is not normalized, see \cite{Albantakis2022_4_0}).
Within system $S$, $\varphi_e(m)$ is then defined as the integrated effect information of $m$ evaluated across all possible purviews $Z \subseteq S$ with $\displaystyle\varphi_e(m) = \max_Z \varphi_e(m, Z)$. 

The effect purview 

\begin{equation}
\label{eq:Zstar}
Z_e^*(m) = \argmax_{Z \subseteq S} \varphi_e(m, Z), 
\end{equation}
in state 

\begin{equation}
z_e^*(m) = \argmax_{\{z_e' | Z \subseteq S\}} \varphi(m, Z= z'_e) = \argmax_{\{z_e' | Z \subseteq S\}} \left( \pi_e(z'_e\mid m) \log \left( \frac{\pi_e(z'_e \mid m)}{\pi_e^{\p'}(z'_e \mid m)} \right) \right)
\end{equation}
corresponds to the maximally irreducible intrinsic effect of $M = m$ within $S$. 

To summarize, 

\begin{equation}
\label{eq:phim}
\varphi_e(m) = \varphi(m, z^*_e) = \max_{Z \subseteq S} \; \left( \pi_e(z'_e\mid m) \log \left( \frac{\pi_e(z'_e\mid m)}{\pi_e^{\p'}(z'_e\mid m)} \right) \right),
\end{equation}
with $\p'$ as in \eqref{eq:mip} and analogously for $\varphi_c(m)$.

Finally, the set of all irreducible causes and effects $\{z_{c/e}^* : m \subseteq s, \; \varphi_{c/e}(m) > 0\}$ within a system $S$ in state $s$ forms the basis of the system's state-dependent cause-effect structure. 

(While the value $\varphi_e(m)$ is unique, there may be multiple purviews $Z^*_e$, or multiple states $z^*_e$ within a purview $Z^*_e$ that maximize $\varphi_e(m)$ \cite{Krohn2017, Moon2019, Barbosa2021, Albantakis2022_4_0}. As outlined in IIT 4.0 \cite{Albantakis2022_4_0}, such ties in $z_e^*$ are resolved according to the ``maximum existence principle'' at the system level by selecting the $z_e^*$ that maximizes the amount of structured information $\Phi$ within the system. 
Here, we apply the simplified criterion that larger purviews are selected in case of ties across purviews with different numbers of units $|Z_e|$, as larger purviews typically allow for larger $\Phi$ values. Any remaining ties are reported in the examples below.)

\subsection{Quantum Systems}

Our objective is to define a quantum version of IIT's mechanism integrated information $\varphi(m)$ that is applicable to composite quantum systems and coincides with the classical measure \cite{Barbosa2021, Albantakis2022_4_0} if there is a classical analog to the quantum system.
To that end, we start with a composite quantum system $Q$ in state $\rho = \sum_s \ket{\psi_s}\bra{\psi_s}$, which can be pure or mixed and is described by its density matrix \cite{Zanardi2018, Kleiner2021}.

$Q$ consists of $n$ units $\H_1, \dots, \H_n$, which are each described by a finite dimensional Hilbert space such that $\H_Q = \bigotimes_{i=1}^n \H_i$. Without lack of generality \cite{Gomez2021}, we will focus on systems constituted of $n$ qubits. 
The system's time evolution is defined by a completely positive (trace-preserving) linear map $\T = \{T_\alpha\}$ \cite{Vedral2002}, as

\begin{equation}
\label{eq:trans}
    \rho_{t+1} = \T(\rho_t) =  \sum_\alpha T_\alpha \rho_t T_\alpha ^\dag.
\end{equation}
%
We will mainly consider unitary transformations ($U$)

\begin{equation}
\label{eq:unitary}
    \rho_{t+1} = U \rho_t U^\dag, 
\end{equation}
%
%
where $U^\dag U = 1$, which means that $U$ is reversible and the inverse of $U$ corresponds to its adjoint ($U^{-1} = U^\dag$). However, we will also address quantum measurements $\mathcal{F} = \{F_\alpha\}$ with $\sum_\alpha F_\alpha^\dag F_\alpha = I$, where the probability of obtaining the result $\alpha$ is given by $\P(\alpha) = tr(F_\alpha^\dag F_\alpha \rho_t)$ in the discussion section.
If $Q$ is an open system with environment $E$, such that the joint system evolves under a unitary transformation, we can evaluate the subsystem $Q$ by treating the environment $E$ in its current state $e_t$ as a fixed background condition (but see Section \ref{limitations} below).  

A mechanism $M \subseteq Q$ is a subset of $Q$ with current state $m = \rho^M_t = tr_{M'}(\rho_t)$ within the corresponding Hilbert space $\H_M = \bigotimes_{i \in M} \H_i$, where $M' = Q \setminus M$ and $tr_{M'}$ denotes the trace over the Hilbert space $\H_{M'}$. 

The quantum integrated information of a mechanism $M$ should quantify how much the state $\rho^M_t$ constrains the state of a purview, a system subset $Z_{t \pm 1} \subseteq Q$, before or after an update $\T$ of the system, compared to a partition $\p$ of the mechanism and purview into $k$ independent parts (Eqn. \eqref{eq:parts}). As above, we will omit the update indices $(t-1, t, t+1)$ unless necessary and focus on effects.

\subsubsection{Quantum cause and effect repertoires}

To translate the cause and effect repertoires into a density matrix description, we first treat the special case of a single purview node $Z = Z_i$ with $|Z| = 1$, for which $\pi_e(Z|m) = p(Z_{t+1}|m_t)$ in the classical case. Replacing the probability distributions with the corresponding density matrices, we obtain 

\begin{equation}
\label{eq:q_pi_e_single}
    \pi_e(Z_i|m) = \rho^{Z_i|m}_{t+1} = tr_{Z_i'}\left(\T(\rho^M \otimes \rmm^{M'})\right), 
\end{equation}
where ' denotes the complement of a set in $Q$ and $\rmm^{M'}$ is the maximally mixed state of $M' = Q \setminus M$ (see also \cite{Kleiner2021, Zanardi2018}). 

Next, we consider the case of purviews comprised of multiple units ($|Z| > 1$). In the classical case, units in $M'$ may induce correlations between units in $Z$, as shown in Figure \ref{fig:cmarg} by example of the COPY-XOR gate. The quantum equivalent of a classical COPY-XOR gate is the CNOT gate (Figure \ref{fig:qCNOT}). For classical inputs, the CNOT behaves identically to the COPY-XOR gate and thus the same considerations apply. This means that, also in quantum systems, extraneous correlations should be discounted when evaluating the causal constraints of a system subset $M$, since they do not correspond to constraints due to the mechanism $M$ itself. In the following, we will use $\rho^{Z|m}_{t+1}$ to denote $tr_{Z'}\left(\T(\rho^M \otimes \rmm^{M'})\right)$, while $\pi_e(Z|m)$ corresponds to the final effect repertoire, after discounting extraneous correlations.

In the quantum case, units in $Z$ may be correlated due to entanglement, which means quantum systems may violate the conditional independence assumption imposed for classical systems (Eqn. \ref{eq:condind}). (Note that incomplete knowledge or a coarse-grained temporal scale can lead to a violation of conditional independence in a classical system, but those ``instantaneous interactions'' are not considered intrinsic to the system and are thus ignored in IIT's causal analysis \cite{Marshall2018}).
Simply inserting Eqn. \eqref{eq:q_pi_e_single} into Eqn. \eqref{eq:effect_repertoire} would inadvertently destroy correlations in $Z$ that are due to entanglement (either preserved or produced during the transformation $\T$). 
In order to correctly capture correlations due to entanglement and discount extraneous correlations due to correlated ``noise'' from units in $M'$, the entanglement structure of $\rho^{Z|m}_{t+1}$ must be taken into account. 

The multipartite entanglement structure of an n-qubit \textit{pure} state $\ket{\psi}$ can be identified through partial traces. 
Following \cite{Zhou2019}, we define a partition $\mathcal{P}^r(V) = \{V^{(1)}, \dots, V^{(r)}\}$ with $r = |\mathcal{P}^r| \leq n$, $\bigcup V^{(i)} = V$ and $V^{(i)} \bigcap V^{(j)} = \emptyset$ if $i \neq j$.

\begin{Definition}
An n-qubit pure state $\ket{\psi}$ is $\mathcal{P}^r$-separable iff it can be written as $\ket{\psi} = \bigotimes_{i=1}^r \ket{\psi^{(i)}}$.
\end{Definition}

In the general case that $\rho^{Z|m}_{t+1}$ is a mixed state, it has to be decomposed into a convex mixture of pure states to identify its entanglement structure. 

\begin{Definition}
\label{def:mixed}
An n-qubit mixed state $\rho$ is $\mathcal{P}^r$-separable iff it can be decomposed into a convex mixture $\rho = \sum_s p_s \ket{\psi_s}\bra{\psi_s}$, with $p_s \geq 0$, $\forall s$ and $\sum_s p_s = 1$, such that every $\ket{\psi_s}$ in the mixture is a $\mathcal{P}^r$-separable pure state $\ket{\psi_s} = \bigotimes_{i=1}^r \ket{\psi_s^{(i)}}$ under the same partition $\mathcal{P}^r$. 
\end{Definition}
Note that Definition \eqref{def:mixed} differs from that in \cite{Zhou2019}, as we require the same partition $\mathcal{P}^r$ for all $\ket{\psi_s}$ in the mixture.

\begin{Definition}
\label{def:entanglement_partition}
Out of the set of partitions $\{\mathcal{P}^r\}_\rho = \{\mathcal{P}^r\vert \rho \text{ is }\mathcal{P}^r\text{-separable}\}$, we define the maximal partition $\mathcal{P}^*(\rho)$ as the one with the maximal number of parts $r^* = \max_{\mathcal{P}^r} r$ and $r^* = |\mathcal{P}^*| \leq n$.
\end{Definition}

\begin{Definition}
\label{def:q_effect_repertoire}
Given the maximal partition $\mathcal{P^*}$ of $\rho^{Z|m}_{t+1}$, we can define the quantum effect repertoire of mechanism $m$ over purview $Z$ as

\begin{equation}
\label{eq:q_effect_repertoire}
\pi_e(Z \mid m) = \bigotimes_{i=1}^{r^*} \pi_e(Z^{(i)} \mid m) = \bigotimes_{i=1}^{r^*} \rho^{Z^{(i)}|m}_{t+1}.
\end{equation}
\end{Definition}
The product in \eqref{eq:q_effect_repertoire} is thus taken over the reduced density matrices of all subsets $Z^{(i)} \subseteq Z$ that are entangled within themselves but not entangled with the other qubits in $Z$.
Note that $\mathcal{P}^*$ is a simple set partition, and should not be confused with the disintegrating partitions $\Part(M,Z)$ \eqref{eq:partitions} used to evaluate the integrated information $\varphi(m,Z,\p)$. Identifying the entanglement structure for multipartite mixed states remains an area of active research \cite{Guhne2009, Li2018, Skorobagatko2021}. For 2-qubit mixed states, separability can be determined using the Peres-Horodecki criterion of the positive partial transform \cite{Peres1996, Horodecki1996}. In general, however, this criterion is only a necessary condition for separability \cite{Horodecki1996} and may thus miss certain complex forms of entanglement \cite{Bennett1999}. 

Several implications follow from the definition of the effect repertoire \eqref{eq:q_effect_repertoire}:
\begin{enumerate}
    \item If $\rho^{Z|m}_{t+1}$ corresponds to a pure state, the purview qubits are fully determined by the mechanism qubits. Thus, $\rho^{Z|m}_{t+1}$ is not influenced by qubits outside of $m$. It follows that $\pi_e(Z|m) = \rho^{Z|m}_{t+1}$ if the latter is pure. This is analogous to the classical case, where $\pi_e(Z|m) = p(Z_{t+1}|m_t)$ if $p(Z_{t+1}|m_t)$ is deterministic. 
    \item Conceptually, entangled subsets are treated as indivisible units in the effect repertoire. If a purview is fully entangled then $\pi_e(Z|m) = \rho^{Z|m}_{t+1}$. 
    \item Extraneous classical correlations are successfully discounted, which means they will not contribute to the integrated information of a mechanism (Figure \ref{fig:qCNOT}).
\end{enumerate}


The cause repertoire of a mechanism in state $m$ over a purview $Z$ also requires causal marginalization (independent noise applied to conditionally independent subsets) to isolate the causal constraints of $m$ over $Z$. In the classical case, the cause repertoire is obtained by applying Bayes' rule to the effect product probabilities. The quantum case is more complex as the entanglement structure of $\rho^M$ might need to be taken into account.

If $\T$ is a unitary transformation \eqref{eq:unitary}, the cause repertoire for any subset $m^{(i)} \in \mathcal{P}^*(\rho^M)$ that is, itself, mutually entangled (e.g. the subset could consist of an entangled pair of qubits) but is not entangled with units of other subsets (e.g. other qubits) can be obtained by applying the adjoint operator $\T^\dag$

\begin{equation}
\label{eq:q_single_cause_repertoire}
    \pi_c(Z \mid m^{(i)}) = \rho^{Z|m^{(i)}}_{t-1} = tr_{Z'}\left(\T^\dag(\rho^{M^{(i)}} \otimes \rmm^{M'^{(i)}})\right).
\end{equation}
\begin{Definition}
\label{def:q_cause_repertoire}
Given the maximal partition $\mathcal{P^*}$ of $\rho^{M}$, we can define the quantum cause repertoire of mechanism $m$ over purview $Z$ as

\begin{equation}
\label{eq:q_cause_repertoire}
\pi_c(Z \mid m) = \frac{\prod_{i=1}^{r^*} \pi_c(Z \mid m^{(i)})}{tr\left(\prod_{i=1}^{r^*} \pi_c(Z \mid m^{(i)}) \right)}.
\end{equation}
\end{Definition}
%
Note that the product here is over parts of $\rho^M$, not of $\rho^{Z|m}_{t-1}$. This introduces an asymmetry in the formulation of cause and effect repertoires, as in the classical case \cite{Oizumi2014, Albantakis2019}. This asymmetry is a direct implication of treating non-entangled subsets as ``physical'' causal units, rather than abstract statistical variables. Causal units are conditionally independent in the present given the past, but not vice versa. This means that in the effect repertoire, purview subsets that are not entangled with other units are conditionally independent given the mechanism and independent noise from outside the mechanism (due to causal marginalization). By contrast, the cause repertoire is inferred from the conditionally independent mechanism subsets, but is not itself conditionally independent. The set of effects specified by a quantum state $\rho_t$ undergoing a unitary transformation ($U$) may thus differ from the set of causes specified by $\rho_{t+1} = U \rho_t U^\dag$ (Figure \ref{fig:qCNOT}). (The assumption of conditional independence, paired with causal marginalization, distinguishes IIT's causal analysis from standard information-theoretical analyses of information flow \cite{Ay2008, Albantakis2019}.)

As pointed out in \cite{Kleiner2021}, the quantum IIT formalism proposed by Zanardi et al. \cite{Zanardi2018} does not include causal marginalization (which was formulated in terms of ``virtual units'' in \cite{Oizumi2014}). We will show below that causal marginalization (\ref{eq:q_effect_repertoire}, \ref{eq:q_cause_repertoire}) is necessary to isolate the causes and effects of system subsets also in the quantum case---an observation that should be of relevance to the causal analysis of quantum systems beyond IIT.

\subsubsection{Quantum intrinsic information (QID)}

Our goal is to define a quantum version of the intrinsic difference measure, which coincides with the classical measure \eqref{eq:ID} \cite{Barbosa2021} in the classical case. 
In quantum information theory, the classical definition of the KLD \eqref{eq:KLD}, or relative entropy, is extended from probability distributions to density matrices based on the von Neumann entropy. The quantum relative entropy of the density matrix $\rho$ with respect to another density matrix $\sigma$ is then defined as:

\begin{equation}
   S(\rho || \sigma) = \Tr \rho \log \rho - \Tr \rho \log \sigma,
\end{equation}
which coincides with the classical case if $\rho \sigma = \sigma \rho$. Unitary operations, including a change of basis, leave $S(\rho || \sigma)$ invariant \cite{Vedral2002}. 
Specifically, if $\rho$ and $\sigma$ are expressed as orthonormal decompositions $\rho = \sum_i p_i \ket{i}\bra{i}$ and $\sigma = \sum_j q_j \ket{j}\bra{j}$, we can write \cite{NielsonChuang2011}

\begin{equation}
   S(\rho || \sigma) = \sum_i p_i \left( \log (p_i) - \sum_j P_{ij} \log (q_j) \right),
\end{equation}
where $P_{ij} = \braket{i}{j} \braket{j}{i}$. In this formulation, a quantum version of the intrinsic difference measure can be defined as

\begin{equation}
\label{eq:QID}
   \QID(\rho || \sigma) = \max_i  p_i \left( \log (p_i) - \sum_j P_{ij} \log (q_j) \right),
\end{equation}
%
analogous to the classical measure. As for the relative entropy, $\QID(\rho || \sigma)$ coincides with the classical case if $\rho \sigma = \sigma \rho$, because in that case $P_{ij} = \delta_{ij}$. Moreover, $\QID(\rho || \sigma) = S(\rho || \sigma)$ if $\rho$ is pure, as in the classical case for fully selective distributions. Otherwise, the QID is subadditive, as desired \cite{Barbosa2020}.

Zanardi et al. \cite{Zanardi2018} proposed the trace distance as a measure of the cause/effect information based on its simplicity and widespread use in quantum-information theory. The trace distance quantifies the maximal difference in probability between two quantum states across all possible POVM measures \cite{NielsonChuang2011}, which is a useful quantity from the perspective of an experimenter. 
By contrast, QID is a measure of the \textit{intrinsic} information of a quantum mechanism. Its value is maximized over the eigenvectors $\{\ket{i}\}$ of $\rho$ \eqref{eq:QID}. 
If $\rho$ is pure, there is only one non-zero eigenvalue and the state identified by the $\QID$ measure is simply $\rho$. If $\rho$ is mixed, the eigenvalue $p_i$ that maximizes equation \eqref{eq:QID} may be degenerate. In that case the QID specifies the eigenspace spanned by the set of eigenvectors for which the difference between $\rho$ and $\sigma$ is maximal. Otherwise, the QID specifies the eigenvector of $\rho$ with the optimal eigenvalue.

\subsubsection{Identifying intrinsic causes and effects}
\label{intrinsicce}
Equipped with the quantum intrinsic difference (QID) measure \eqref{eq:QID}, the intrinsic effect information that the quantum mechanism $M = m$ specifies over a purview $Z$ can be quantified as

\begin{equation}
    \ii_e(m,Z) = \QID\left(\pi_e(Z|m), \pi_e(Z)\right),
\end{equation}
where $\pi_e(Z) = \pi_c(Z) = \rho^Z_{mm}$ is the maximally mixed state in the quantum case.

Following from equation \eqref{eq:QID}, with $\rho = \pi_e(Z\mid m) = \sum_i p_i \ket{i}\bra{i}$ as the effect repertoire and $\sigma = \pi_e(Z) = \sum_j q_j \ket{j}\bra{j} = \rho^Z_{mm}$ as the unconstrained effect repertoire, the intrinsic effect of mechanism $m$ on purview $Z$ is

\begin{equation}
\begin{split}
\label{eq:z'}
    z_e'(m,Z) &= \argmax_{i \in \mathcal{H}_{Z}} p_i \left( \log p_i - \sum_j P_{ij} \log (q_j) \right) \\
    &= \argmax_{i \in \mathcal{H}_{Z}} p_i \left( \log p_i - \log |\mathcal{H}_{Z}|^{-1} \right),
\end{split}
\end{equation}
where $|\mathcal{H}_{Z}|$ denotes the cardinality of $\mathcal{H}_{Z}$. 
The intrinsic effect $z'_e(m,Z)$ is thus simply the eigenvector $\ket{i}$ of $\pi_e(Z|m)$ with the maximal eigenvalue. If the maximal eigenvalue of $\rho = \pi_e(Z\mid m)$ is degenerate, $z^*_e(m)$ corresponds to the subspace of $\mathcal{H}_{Z_e^*}$ spanned by the set of eigenvectors belonging to the maximal eigenvalue
(and the same for the intrinsic cause $z'_c(m,Z)$ evaluated over $\pi_c(Z|m)$).

Note that, in the case that $\pi_e(Z|m)$ is a mixed quantum state (corresponding to a probability distribution with multiple possible effect states in the classical case), this means that the \textit{intrinsic} effect $z'_e(m,Z)$ differs from $\rho = \pi_e(Z|m) = \sum_i p_i \ket{i}\bra{i}$.

\subsubsection{Disintegrating partitions}

As in the classical case, the quantum integrated information $\varphi(m,Z,\theta)$ is evaluated by comparing the effect repertoire $\pi_e(Z|m)$ to a partitioned effect repertoire $\pi^\theta_e(Z|m)$ (and analogously for $\varphi_c(m, Z, \theta)$).

The set of possible partitions $\p \in \Part(M,Z)$ is the same as for the classical case (Eqn. \ref{eq:partitions}). Likewise, the partitioned effect repertoire is defined as in \eqref{eq:partitioned_effect}, as a product over the parts in the partition. 
In the quantum case, $\pi_e(Z^{(i)}|\varnothing)$ corresponds to the maximally mixed state $\rmm^{Z^{(i)}}$. 
The partitioned cause repertoire is defined in the same way.

Note that the disintegrating partition $\theta \in \Theta(M,Z)$ \eqref{eq:partitions} here is applied on top of $\mathcal{P}^*$ (Definition \ref{def:entanglement_partition}). Partitioning may thus affect entanglement within the repertoire. 
Conceptually, any entanglement in $\pi_e(Z\mid m)$ that is destroyed by the partition $\p$ will count towards $\varphi_e(m, Z, \p)$. Ultimately, however, $\varphi_e(m, Z)$ is again evaluated over $\p'$ \eqref{eq:mip}, the minimum information partition (MIP). This means that everything else being equal, partitions that affect entanglement less are more likely to correspond to the MIP.

\subsubsection{Quantum mechanism integrated information}

Having identified the specific effect state $z'_e$ as an eigenstate $\ket{i}$ of $\rho = \pi_e(Z|m)$, the integrated effect information $\varphi(m, Z, \theta)$ is evaluated as the $\QID(\rho||\sigma)$ over that eigenstate, such that

\begin{equation}
\label{eq:qphi}
    \varphi(m,Z,\theta) = \varphi(m, z'_e, \theta) = p_i \left( \log p_i - \sum_j P_{ij} \log (p^\theta_j) \right),
\end{equation}
where $\sigma = \pi^\theta_e(Z|m) = \sum_j p^\theta_j \ket{j}\bra{j}$ is now the partitioned effect repertoire.

As above, quantifying the integrated effect information $\varphi_e(m)$ of a mechanism $m$ within a quantum system $Q$ requires a search over all possible partitions $\p \in \Part(M,Z)$ to identify the MIP, and a search across all possible purviews $Z \subseteq Q$, such that

\begin{equation}
 \varphi_e(m) = \max_{Z \subseteq Q} \varphi_e(m, Z) = \max_{Z \subseteq Q} \varphi(m, Z, \theta'),   
\end{equation}
as in \eqref{eq:phim}, with $\theta'$ as in \eqref{eq:mip}, and analogously for $\varphi_c(m)$. 

The maximally irreducible effect purview $Z^*_e(m)$ 

\begin{equation}
    Z^*_e(m) = \argmax_{Z \subseteq Q} \varphi_e(m, Z)     
\end{equation}
again corresponds to the subset of $Q$ upon which the mechanism $M = m$ has the maximally irreducible intrinsic effect $z^*_e$, which corresponds to the eigenstate of $\rho = \pi_e(Z^*|m)$ that maximizes Eq. \eqref{eq:z'}, or the eigenspace spanned by a set of eigenvectors corresponding to a degenerate maximal eigenvalue.

As in the classical case, $Z^*_e$ is not necessarily unique and we again choose the larger purview in case of a tie between purviews of different sizes (see above). Any remaining ties are reported in the examples below.

\subsubsection{The intrinsic structure of a quantum system}

Standard approaches for studying the causal or informational properties of a system typically assume either a reductionist perspective (focused on individual units) or holistic perspective (describing the system as a whole). 
As the units in a quantum system can be entangled, focusing on individual units is ill-suited at the quantum level. However, a purely holistic description of a quantum system will still miss differences in the internal structure of a quantum state (see below the comparison between the maximally entangled GHZ-type and W-type states \cite{Dur2000}).

In IIT, causation is neither reductionist nor holistic but compositional: the IIT analysis considers the intrinsic causes and effects of every subset within a system and quantifies their irreducibility as $\varphi_{c/e}(m)$ \cite{Albantakis2019cc}. As a result, it can elucidate the internal structure of composite quantum states and operators, as we will show in the next section.

We note that typically, the IIT analysis assumes a current system state $s_t$ and identifies its compositional causes at $t-1$ and effects at $t+1$. A subset $m \subseteq s$ with an irreducible cause and effect forms a ``causal distinction'' within the system $s$, where $\varphi(m) = \min(\varphi_c(m), \varphi_e(m))$ is the integrated (cause-effect) information of $m$. 

According to IIT, the phenomenal experience of a physical system $S$ in state $s$ is identical to its cause-effect structure, composed of a system's causal distinctions and their relations \cite{Haun2019}. 
Unfolding the full cause-effect structure requires assessing the integrated (cause-effect) information $\varphi(m)$ of every subset of units $m \subseteq s$.

For ease of demonstration, in the following, we will instead evaluate examples of system transitions from state $t$ to $t + 1$ and identify the intrinsic effects of the system in state $s_t$ and the intrinsic causes of the system in state $s_{t+1}$ (see also \cite{Albantakis2019}). 

\section{Results}

For a direct comparison between classical and quantum systems, we will focus our attention on computational quantum systems (see \cite{Mermin2007} for an overview and comparison to classical systems), constituted of a finite number of quantum units with a finite-dimensional Hilbert space, evolving in discrete updates according to unitary transformations, expressed in the computational (or “classical”) basis unless stated otherwise.

To compute classical IIT quantities, we made use of the openly available PyPhi python toolbox, developed by the Tononi lab \cite{Mayner2018, Gomez2021}, using the ``iit-4.0'' feature branch with standard IIT 4.0 settings. To compute quantum IIT results, we implemented a QIIT toolbox (\url{https://github.com/Albantakis/QIIT}, accessed 30 December 2022), applicable to unitary quantum mechanisms of two and three qubits.

\subsection{CNOT}

\subsubsection{Classical case} 
As a first example, we will evaluate the ``controlled-NOT'' (CNOT) gate. Classically, the CNOT gate corresponds to a reversible XOR gate, with a COPY operation performed on the first input bit (A) and an XOR operation comparing the two input bits A and B as the second output (Figure \ref{fig:cCNOT}). For instance, the input state $AB = (1,0)$ leads to the output $CD = (1,1)$. In what follows, we will abbreviate the states of system subsets (mechanisms and purviews) by the state plus a set subscript, for example, $10_{AB}$ for $AB = (1,0)$. 

Given the input state $AB = (1,0)$, the IIT analysis identifies two irreducible mechanisms, one first-order and one second-order mechanism. The mechanism $1_A$ specifies the effect purview $1_C$ with $\varphi = 1 \text{ ibit}$; the second-order mechanism $10_{AB}$ specifies the effect purview $11_{CD}$ also with $\varphi = 1 \text{ ibit}$ (while there is a tie with the effect $1_{D}$, we choose the larger purview as described above). Notably, $0_B$ by itself (with A replaced by independent noise) does not specify any information about the next state of CD (Figure \ref{fig:cmarg}). While this conclusion should be straightforward, it relies on the use of product probabilities instead of simple conditional probabilities \eqref{eq:effect_repertoire}. The latter would mistakenly count the correlation between C and D as an effect of B, although it is actually due to the common input A. 

By contrast to $0_B$ on the effect side, $1_D$ on the cause side specifies irreducible cause information about the previous state of $AB$ in addition to $1_C$ and $11_{CD}$, albeit only $\varphi_c(1_D) = 0.5$ ibit due to the remaining uncertainty about the state of $AB$ (note the quantitative difference between the ID measure \eqref{eq:ID} and the KLD \eqref{eq:KLD}, which would return a value of 1 bit).

\subsubsection{Quantum case}
\begin{figure}[b!]
    \includegraphics[width=\columnwidth]{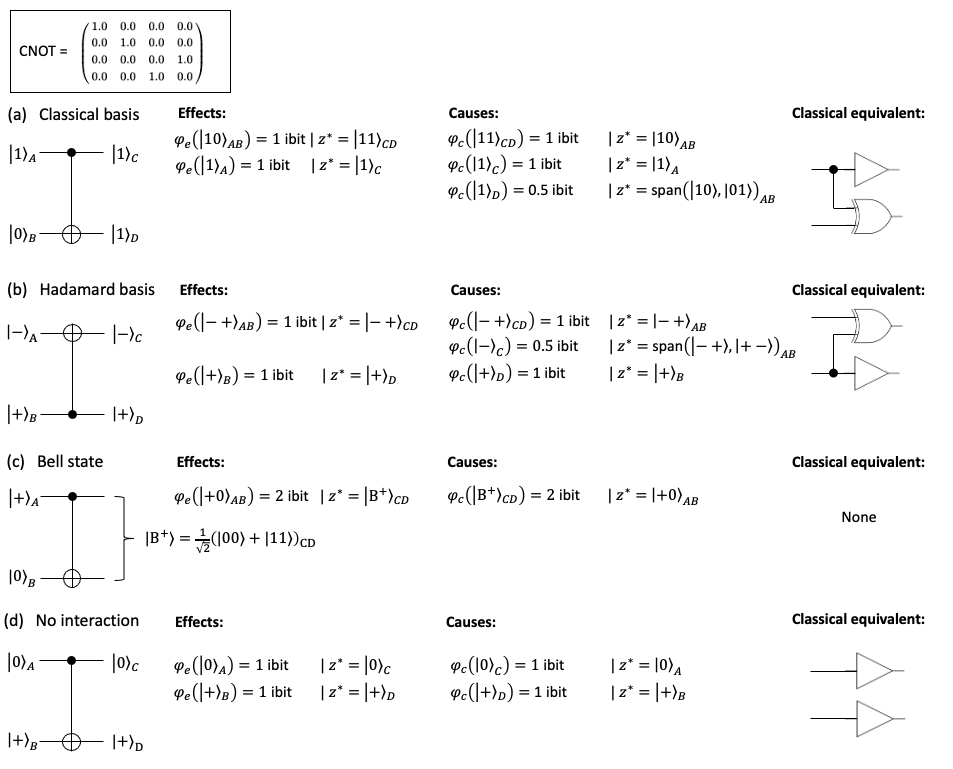}
    \caption{CNOT gate. The CNOT operator is shown in the top box. \textbf{(a)} For a pure input state in the classical basis, we obtain the same results as in the classical case (Fig. \ref{fig:cCNOT}). \textbf{(b)} For a pure input state in the Hadamard basis, the role of the ``control'' (here $B$) and ``target'' (here $A$) is reversed compared to \textbf{(a)} (as indicated in the circuit diagram). \textbf{(c)} The CNOT is often used to produce a ``Bell state'' of two maximally entangled qubits. In this exclusively quantum scenario, only the second order mechanisms $\ket{+0}_{AB}$ and $\ket{B^+}_{CD}$ specify an effect or cause, respectively. None of the subsets has any cause or effect information ($\varphi = 0$ ibit). \textbf{(d)} Conversely, given the input state $\ket{0+}_{AB}$ all second order mechanisms are fully reducible ($\varphi = 0$ ibit) and only the first order mechanisms specify causes and effects.}
    \label{fig:qCNOT}
\end{figure}

For a CNOT gate with the input state $\rho^{\text{AB}} = \ket{10}\bra{10}$ (or $\ket{10}_{AB}$), we obtain the same results as for a COPY-XOR gate with input state $AB=(1,0)$ using the formalism outlined above (Figure \ref{fig:qCNOT}a). With essentially classical inputs, the CNOT gate thus reproduces the intrinsic causal structure of the classical COPY-XOR gate. 

To that end, it was necessary to discount the spurious correlation between qubits $A$ and $B$ through product distributions \eqref{eq:q_effect_repertoire}. This demonstrates that standard conditional probabilities are insufficient to identify the causes and effects of system subsets also in the quantum case.

Note that for the CNOT gate the role of the ``control'' (COPY) and the ``target'' (XOR) qubit changes depending on the input state, which is not true for the COPY-XOR gate. For an input state in the Hadamard basis, e.g. $\ket{-+}_{AB}$, information seems to flow from B to C, not A to D as for a classical input. Accordingly, the quantum IIT analysis now identifies $\ket{+}_B$ and $\ket{-+}_{AB}$ as irreducible mechanisms with $\varphi = 1 \text{ ibit}$, while $\ket{-}_A$ by itself does not specify any effect information (Figure \ref{fig:qCNOT}b). Yet, $\ket{+}_C$ does specify irreducible cause information about AB. 

In quantum systems, the CNOT is often used to produce the maximally entangled Bell state $\ket{B^+} = \frac{1}{\sqrt{2}} (\ket{00} + \ket{11})$. $CD = \ket{B^+}$ results from the input state $AB = \ket{+0}$, a transition for which there is no classical circuit equivalent \cite{Schumacher2012}. In this case, the quantum IIT analysis identifies only the second order mechanisms (constituted of two qubits) $\ket{+0}_{AB}$ and $\ket{B^+}_{CD}$ with $\varphi = 2$ ibit each. Individual qubits specify no cause or effect information (Figure \ref{fig:qCNOT}c). An analogous result obtains for the Bell state as the input to the CNOT gate.

Finally, with $AB = \ket{0+}$ as the input, there appears to be no interaction between qubits and the quantum IIT analysis only identifies first order mechanisms on the cause and effect side (Figure \ref{fig:qCNOT}d).

\subsubsection{Mixed states and extensions to larger systems}

The purpose of the IIT analysis is to evaluate the cause-effect power of a system in its current state. Evaluating statistical ensembles is conceptually not in line with the theory. Accordingly, the classical IIT analysis always assumes a particular (fully determined) state for the mechanism $m$. However, in quantum mechanics, mixed states not only describe statistical ensembles, but also subsets of entangled pure states.

\begin{figure}[b!]
    \includegraphics[width=\columnwidth]{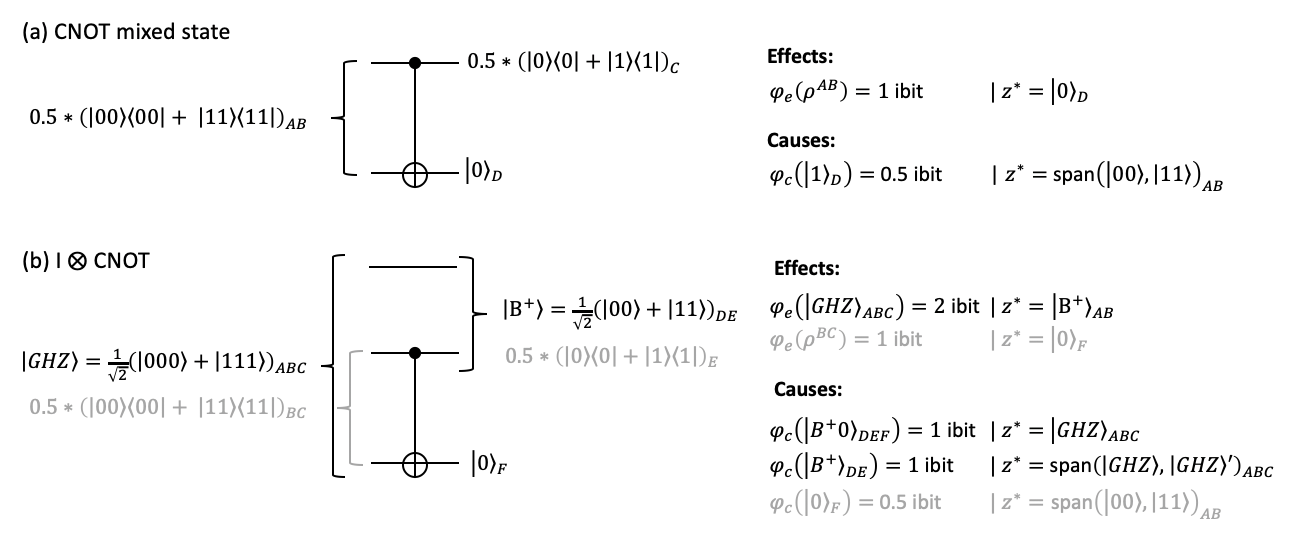}
    \caption{Mixed states and entanglement with the environment. \textbf{(a)} IIT analysis of the CNOT gate with a mixed input state $\rho^{AB} = 0.5*(\ket{00}\bra{00}+\ket{11}\bra{11})$. \textbf{(b)} It is possible to describe the mixed state as a pure state entangled with the environment. Analyzing such an extended system for the case in (a), the cause and effect of the subsystem are preserved in the larger system (gray), but we obtain additional causes and effects that span all three qubits (black). $\ket{GHZ}'$ denotes a maximally entangled superposition of states $\ket{001}$ and $\ket{110}$.}
    \label{fig:qCNOTmixed}
\end{figure}

If we apply an even mixture $\rho^{AB} = 0.5*(\ket{00}\bra{00}+\ket{11}\bra{11})$ to the CNOT gate, we obtain $\rho^{CD} = 0.5*(\ket{00}\bra{00}+\ket{10}\bra{10})$ as a result. In this case, only the second order mechanism $m = \rho^{AB}$ has an irreducible effect with $\varphi_e = 1.0$ ibit over $z^* = \ket{0}_D$. There is no effect on C, as C by itself is undetermined (maximally mixed). In turn, only $\ket{0}_D$ specifies an irreducible cause with $\varphi(\ket{0_D}) = 0.5$ ibit over purview $Z^* = AB$ with $z^*$ corresponding to the subspace spanned by $\ket{00}_{AB}$ and $\ket{11}_{AB}$ (Figure \ref{fig:qCNOTmixed}a). 
Note the difference to the causal analysis of the Bell state $\ket{B^+} = \frac{1}{\sqrt{2}} (\ket{00} + \ket{11})$ above, where $\ket{+0}_{AB}$ and $\ket{B^+}_{CD}$ both specified second order mechanisms with $\varphi = 2$ ibit each.

The same transition may also be described as part of a larger system of three qubits. To that end, we can extend the CNOT gate by an identity operator acting on the additional qubit (Figure \ref{fig:qCNOTmixed}b), which may stand for the environment.
Assuming that the three qubits (ABC) are initially in a maximally entangled GHZ state \cite{Greenberger1989}, the state after applying I$\otimes$CNOT leaves the first two qubits (DE) maximally entangled, while the third qubit (F) is in state $\ket{0}_F$. The causal analysis of the 3 qubit system reveals additional causes and effects that span all three qubits, but also includes the cause and effect identified for the mixed 2 qubit subsystem evaluated in Figure \ref{fig:qCNOTmixed}a.

\subsubsection{Intrinsic structure due to entanglement}

The IIT analysis evaluates the potential causes and effects of a system in a state before and after an update of the system \eqref{eq:trans}. In the classical case, there is no instantaneous interaction between the units of a system (which corresponds to the conditional independence assumption \eqref{eq:condind} \cite{Albantakis2019}). In the quantum case, however, entanglement between qubits can lead to additional intrinsic structure (see also \cite{Kremnizer2015}). To identify the intrinsic structure of a quantum state that is due to entanglement, we can assume $\T = I$ (the identity operator) in \eqref{eq:trans}.
In that case, causes and effects are equivalent and should be viewed as constraints of the quantum state onto itself.

\begin{figure}[h!]
    \includegraphics[width=10 cm]{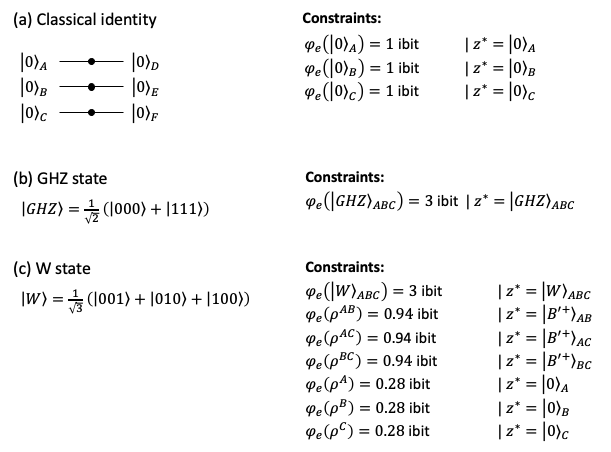}
    \caption{Intrinsic structure of 3-qubit states. \textbf{(a)} Classical states specify first order constraints under an identity function (equivalent to 3 classical COPY gates). \textbf{(b)} The maximally entangled GHZ-state only specifies a 3\textsuperscript{rd} order constraint. \textbf{(c)} By contrast, the W-state, which is also maximally entangled, specifies constraints of all orders. Subsets $m \subseteq s$ of the W-state are indicated by $\rho^m$. The remaining units $s\setminus m$ are traced out. $\ket{B'^+}$ indicates a superposition of $\ket{10}$ and $\ket{01}$.}
    \label{fig:identity}
\end{figure}

For classical states, causal analysis identifies only first order constraints for $\T = I$ (Figure \ref{fig:identity}a).
The entanglement of tripartite quantum states is not a trivial extension of the entanglement of bipartite systems \cite{Acin2001}. In addition to biseparable states (A-BC, B-AC, C-AB), there exist two classes of genuine tripartite entanglement: GHZ-type and W-type states \cite{Dur2000}. For the GHZ-state $\ket{\text{GHZ}} = \frac{1}{\sqrt{2}}(\ket{000}+\ket{111})$, all subsystems correspond to unentangled, evenly mixed states of zeros and ones.
For the W-state $\ket{\text{W}} = \frac{1}{\sqrt{3}}(\ket{001}+\ket{010}+\ket{100})$ all bipartite subsystems remain entangled with different probabilities of zeros and ones. The difference between these two states is clearly identified by the IIT analysis. While the GHZ-state only specifies a 3\textsuperscript{rd} order constraint without any substructure, the W-state has full structure with intrinsic constraints on all subsets.

\section{Discussion}

Our goal in this study was to extend the mathematical formalism of IIT from discrete, classical dynamical systems to finite-dimensional quantum systems, starting with IIT's mechanism integrated information $\varphi(m)$ \cite{Barbosa2021}. To that end, we translated IIT's intrinsic difference measure \cite{Barbosa2020, Barbosa2021} into a density matrix formalism, and extended the notion of conditional independence and causal marginalization \cite{Albantakis2019} to allow for quantum entanglement.
Our results demonstrate that it is possible to extend the applicability of IIT's formal framework to finite-dimensional quantum systems evolving according to unitary transformations, such that the quantum formulation converges to the classical formulation for essentially classical state updates (as demonstrated by the example of the CNOT gate, Figures \ref{fig:cCNOT} and \ref{fig:qCNOT}).
In the following, we will compare our work to previous attempts of applying IIT to quantum systems \cite{Tegmark2015, Zanardi2018, Kleiner2021}, discuss several difficulties in applying IIT's causal analysis to measurement dynamics, and highlight several limitations and implications of our QIIT formalism.


\subsection{Comparison with previous approaches}

Potential extensions of IIT to quantum systems have been explored in \cite{Tegmark2015, Kremnizer2015, Zanardi2018, Kleiner2021, Chalmers2021}. Of these, only Zanardi et al.~\cite{Zanardi2018} aimed for a direct translation of the IIT formalism (specifically, ``IIT 3.0'' \cite{Oizumi2014}) from a classical into a quantum-mechanical framework. 
As demonstrated by Kleiner and Tull \cite{Kleiner2021}, the quantum IIT formalism proposed in \cite{Zanardi2018} captures the higher-level mathematical structure of the canonical framework (IIT 3.0). However, it does not converge to the classical IIT framework and thus does not allow for a quantitative comparison across quantum and classical systems.
Among other differences, Zanardi et al. omitted the causal marginalization of variables outside the cause or effect repertoires and across partitions. As we have shown above (Figure \ref{fig:cCNOT} and \ref{fig:qCNOT}), causal marginalization is necessary to identify the causal constraints specific to a subset of variables within the system also in the quantum case. 
Paired with the conditional independence assumption, this also implies that the IIT formalism does not obey time-reversal symmetry, even when applied to unitary transformations (see also \cite{Albantakis2019cc} for classical reversible systems).

Compared to \cite{Zanardi2018}, we have, moreover, incorporated several updates of the IIT formalism from ``IIT 3.0'' \cite{Oizumi2014} to ``IIT 4.0'' \cite{Barbosa2021, Albantakis2022_4_0}. These include an updated partitioning scheme \cite{Albantakis2019, Barbosa2021}, as well as a novel measure of intrinsic information based on the intrinsic difference (ID) introduced in \cite{Barbosa2020}. While Zanardi et al. \cite{Zanardi2018} used the trace distance to quantify $\varphi$, we have developed a quantum version of the novel intrinsic information measure, starting from the quantum relative entropy between two density matrices. 
In combination with the implementation of causal marginalization in quantum systems, the QIIT formalism proposed above thus converges to the classical version for essentially classical state updates.

While \cite{Zanardi2018, Kleiner2021} are mainly concerned with the mathematical framework of IIT, \cite{Kremnizer2015} and \cite{Chalmers2021} apply the notion of integrated information within the context of a consciousness-induced collapse model of quantum mechanics. To that end, Chalmers and McQueen \cite{Chalmers2021} utilize the QIIT framework proposed in \cite{Zanardi2018}. Kremnitzer and Ranchin \cite{Kremnizer2015} present an independent quantum integrated information measure based on the quantum relative entropy. However, their measure applies to the quantum state itself and does not take the dynamics of the quantum system into account.
Our work has a different focus. IIT does not require a role for consciousness in the collapse of the wave function (but see section \ref{sec:measurements} below).

Finally, Tegmark \cite{Tegmark2015} leans on the general principles of IIT's approach to understand and explain consciousness in physical systems and addresses the so-called ``quantum factorization problem'' \cite{Schwindt2012} using generalized measures of information integration.
While we regard the quantum factorization problem as a serious issue, it is beyond the scope of this work. Our assumed starting point is a particular density matrix that undergoes a particular unitary transformation \eqref{eq:unitary}. While the QID measure \eqref{eq:QID} is basis independent, a system's cause-effect structure and the mechanism integrated information values $\varphi(m)$ of its subsets $m \subseteq Q$ will typically change under an additional unitary transformation.

\subsection{Measurement dynamics}
\label{sec:measurements}

The dynamics of a quantum measurement can be described by a quantum operator $\mathcal{F} = \{F_\alpha\}$ with $\sum_\alpha F_\alpha^\dag F_\alpha = I$. While the output of a unitary transformation is a density matrix corresponding to a pure or mixed quantum state, the outcome of a measurement is probabilistic with $\P(\alpha) = tr(F_\alpha^\dag F_\alpha \rho_t)$ for measurement outcome $\alpha$ \cite{NielsonChuang2011}. 

The IIT analysis evaluates the potential effects and potential causes of a mechanism in a state. From the perspective of the quantum state $\rho_t$ being measured, the measurement outcome is still unknown. The effect repertoire of the quantum state $\rho_t$ directly before the measurement \eqref{eq:q_effect_repertoire} could thus be computed from

\begin{equation}
\label{eq:e_measurement}
    \rho_{t+1} = \sum_\alpha F_\alpha \rho_t F_\alpha^\dag,
\end{equation}
following equation \eqref{eq:trans}. The density matrix $\rho_{t+1}$ then corresponds to a mixed state, that is, a probability distribution of possible measurement outcomes.

Measurement dynamics become problematic, if we want to evaluate the quantum state directly after the measurement. Here, the cause repertoire has to be computed from the perspective of the quantum state post measurement $\rho^{\alpha}_{t+1}$, corresponding to a particular measurement outcome $\alpha$ 

\begin{equation}
\label{eq:c_measurement}
    \rho^{\alpha}_{t+1} =  \frac{F_\alpha \rho_t F_\alpha^\dag}{tr(F_\alpha^\dag F_\alpha \rho)}.
\end{equation}

Since measurements are not unitary transformations, the adjoint operator $\T^\dag$ is not the same as the inverse $\T^{-1}$. For this reason, we cannot use equation \eqref{eq:q_single_cause_repertoire} to compute the cause repertoire of $\rho^{\alpha}_{t+1}$ (note that the same holds for prior proposals \cite{Zanardi2018, Kleiner2021}). 

In the classical case, the cause repertoire of an irreversible mechanism can be computed using Bayes' Rule \cite{Barbosa2021, Albantakis2022_4_0}. However, in the quantum case, all information about the basis of the original quantum state before the measurement is lost, which means that there are infinitely many possible past states. While different past states should still be more or less likely, we do not know of any available method for obtaining a probability distribution of possible causes in this case. 

That said, the amount of cause information specified by a post-measurement state $\rho^{\alpha}_{t+1}$ depends on the way the measurement dynamics are conceptualized, and thus on the specific quantum theory applied. 
While $\rho^{\alpha}_{t+1}$ specifies (almost) no cause information under spontaneous collapse theories, the case may be quite different for deterministic hidden variable theories. 
No, or very low, cause information at the quantum level would imply that quantum systems are poor substrates for consciousness and may offer room for macro level descriptions to reach maximal values of integrated information, as predicted by IIT.

Finally, the technical difficulties introduced by probabilistic measurement dynamics would naturally be avoided by so-called ``no-collapse'' models of quantum mechanics, such as the Many-Worlds Interpretation. However, theories that rely only on a density matrix encoding the state of the universe and a unitary transformation determining its time-evolution \cite{Tegmark2015} face a different issue when it comes to identifying conscious entities through causal, informational, or computational means. If applied at the fundamental level, any entities obtained would correspond to subsets of the universal density matrix, never subsets within individual ``branches'' only (see for example Fig. \ref{fig:qCNOT}c). While the QIIT measures (and other quantities) could formally be applied within a branch, there is no principled justification for doing so from the perspective of a fundamental theory of consciousness (note that the notion of decoherence cannot resolve this issue).  

\subsection{Formal considerations and limitations}
\label{limitations}
Formally, the restriction to unitary transformations eliminated differences between the unconstrained cause and effect repertoire that commonly arise in the classical formulation. Nevertheless, due to the assumption of conditional independence on the effect side, but not the cause side, cause repertoires are formally distinct from effect repertoires even under unitary transformations. 

The quantum formulation also provides justification for treating all variables outside the candidate system under consideration as fixed background conditions, which is motivated by IIT's intrinsicality postulate \cite{Albantakis2019, Oizumi2014}: by the no-communication theorem \cite{NielsonChuang2011}, any unitary transformation on a system will leave the density matrix of its environment unchanged. 
However, not all subsets of unitary transforms are unitary. Future work should explore the implications of assuming fixed background conditions in such cases.

The IIT formalism for classical systems starts from a transition probability matrix (TPM) which corresponds to a complete set of transition probabilities (from every possible system state to every possible system state) \eqref{eq:tpm}. This has led some to critize IIT on conceptual grounds, as it seems to imply that subjective experience would depend not only on the actual states a system inhabits in the course of its dynamical evolution, but also on hypothetical counterfactuals that may never happen \cite{seth2021being}. 
In the QIIT formalism, the role of the classical transition probability matrix (TPM) is assumed by the unitary transform \eqref{eq:unitary} applied to the quantum state. 
Just as evolution operators in quantum mechanics essentially are TPMs, in IIT, the TPM simply serves as a complete description of the system's dynamics. 

In this work, we have focused on mechanism integrated information $\varphi$ \cite{Barbosa2021}. In principle, it should be possible to formally extend our QIIT formalism to incorporate the full ``IIT 4.0'' framework, including the system integrated information ($\varphi_s$) \cite{Marshall2022}, a full characterization of the system's cause-effect structure comprised of causal distinctions and causal relations \cite{Haun2019}, and the amount of structure information ($\varPhi$) specified by a system. 

Nevertheless, there are several conceptual issues that need to be resolved before the QIIT formalism can be applied to identify conscious systems, which have to comply with all of IIT's requirements for being a substrate of consciousness (IIT's ``postulates'') \cite{Oizumi2014}. 
For example, it is unclear whether mixed states should count as permissible states for evaluating the system's integrated information. While only specific sets of units, not ensembles, qualify as substrates, a particular set of units may still be in a mixed state if it is entangled with the environment (Fig. \ref{fig:qCNOTmixed}).
Yet, IIT's information postulate requires systems and mechanisms to have specific cause-effect power. It thus remains to be determined whether mixed states can comply with IIT's information postulate.

Recurrent quantum systems are another issue. In the classical formulation, recurrent connections between system units are required for positive system integrated information \cite{Oizumi2014, Marshall2022}. Physical units (e.g., neurons, transistors,...) are thus assumed to be dynamically persistent variables with at least two possible states. However, it is less obvious whether qubits, or qudits more generally, may indeed be treated as variables that maintain a causal identity across their state updates.

\subsection{From micro to macro?}

Current empirical evidence suggests that consciousness and its contents are correlated with the dynamics and activity of neurons in some parts of the cerebral cortex \cite{Koch2016}. 
While our experiences seem to unfold over macroscopic spatial and temporal scales, the brain can, in principle, be described at a multitude of levels, for example, as a network comprised of a few interacting brain regions, or a microphysical quantum system. Why is it then that the contents of our experiences correlate with neural activity in certain regions of the cortex rather than their underlying microphysical processes \cite{Tegmark2015,Hoel2016}?

IIT offers a single, general principle for identifying conscious systems: a substrate of consciousness must correspond to a set of units that forms a maximum of intrinsic cause-effect power over grains of units, updates, and states \cite{Tononi2016, Hoel2016, Marshall2018}. However, it remains to be determined whether IIT's propositions are compatible with our current best knowledge about micro physics \cite{Barrett2019, Carroll2021}. 

The QIIT formalism presented above allows for a quantitative comparison between classical and quantum systems. Squaring IIT (as well as any other causal, computational, or information-based theory of consciousness) with our current knowledge of micro physics, moreover, requires a method for obtaining macroscopic causal models from microscopic dynamics. This could be achieved by a "black-boxing" of quantum circuits into suitable macro-units \cite{Marshall2018}, or a quantitative framework that formalizes the emergence of well-defined probability distributions \cite{Durham2020}. 

To identify the maximally irreducible description of a system across a hierarchy of spatio-temporal scales, we have to compare micro- and macro-level descriptions of the \textit{same} system. While it is always possible to implement the global function performed by a classical system with a quantum circuit \cite{NielsonChuang2011}, these systems will typically not have the same causal structure (the CNOT gate described in Fig. \ref{fig:qCNOT} is exceptional in that way). One reason is that quantum gates have to be reversible, and thus require so-called ``ancilla qubits'' to implement convergent logic gates such as AND-gates, or NOR-gates. These ancilla qubits cannot simply be ignored in the IIT analysis, as this would introduce an observer-dependent, extrinsic perspective. They also cannot typically be treated as a fixed background conditions. Understanding whether and how irreversible logic functions might emerge from reversible quantum circuits is thus an important subject for future investigations.

As is, QIIT and its classical counterpart are only partially overlapping in their domains of applicability. While QIIT is in principle more fundamental as an extension of IIT's classical causal framework to quantum systems, it is currently limited to reversible, unitary transformations, and thus cannot directly be applied to irreversible processes. 

Overall, we see it as a positive development that the updated IIT 4.0 formalism for computing the mechanism integrated information \cite{Barbosa2021} is readily applicable within a quantum mechanical framework. Our work revealed several conceptual issues regarding theories of consciousness as they relate to fundamental physics. Regardless, the theoretical framework for identifying causes and effects of subsets of units within a quantum system should be of interest within the field of quantum information theory and quantum causal models more generally.

\vspace{6pt} 



\authorcontributions{``Conceptualization, L.A., R.P., and I.D.; methodology, L.A. and I.D.; software, L.A.; validation, L.A., R.P., and I.D.; formal analysis, L.A.; investigation, L.A.; writing---original draft preparation, L.A.; writing---review and editing, R.P. and I.D; visualization, L.A.. All authors have read and agreed to the published version of the manuscript.''}

\funding{L.A. acknowledges the support of a grant from the Templeton World Charity Foundation (TWCF-2020-20526).}

\acknowledgments{The authors thank the Association for Mathematical Consciousness Science (AMCS) for its institutional support. L.A. thanks William Marshall, Alireza Zaeemzadeh, and Giulio Tononi for helpful discussions and comments on earlier drafts of this article, Will Mayner for maintaining PyPhi and his advice on implementing the QIIT toolbox, and Matteo Grasso for his help with figures. R.P. acknowledges the support of the Munich Center for Mathematical Philosophy.}

\conflictsofinterest{The authors declare no conflict of interest. The funders had no role in the design of the study; in the collection, analyses, or interpretation of data; in the writing of the manuscript, or in the decision to publish the~results.} 

\reftitle{References}


\externalbibliography{yes}
\bibliography{IIT_bibtex.bib}

%


\end{document}